\documentclass[english,twoside,a4paper,10pt]{article}

\usepackage[latin1]{inputenc}
\usepackage[T1]{fontenc}
\usepackage{amsmath}
\usepackage{amsfonts}
\usepackage{graphicx}
\usepackage{subfigure}
\usepackage{a4wide}
\usepackage{amssymb}
\usepackage{fancyhdr}
\usepackage{mathrsfs}
\usepackage[toc,page]{appendix}
\usepackage{xcolor}
\usepackage{cite}

\begin{document}

\title{Local Solutions of General Relativity in the Presence of the Trace Anomaly}

\author{ 
Marco Calz\'a$^1$\footnote{E-mail address: mc@student.uc.pt
},\,\,\,
Alessandro Casalino$^{2}$\footnote{E-mail address: alessandro.casalino@unibo.it}
,\,\,\,
Lorenzo Sebastiani$^{3,4}$\footnote{E-mail address: lorenzo.sebastiani@unitn.it}\\
\\
\begin{small}
$^1$ CFisUC, Departamento de Fisica, Universidade de Coimbra, 3004-516 Coimbra,
Portugal.
\end{small}\\
\begin{small}
$^2$
Dipartimento di Fisica ed Astronomia "Augusto Righi", Universit\'a di Bologna,
\end{small}\\
\begin{small}
Via Piero Gobetti 93/2, 40129 Bologna, Italy.
\end{small}\\
\begin{small}
$^3$ Istituto Nazionale di Fisica Nucleare, Sezione di Pisa, Italy.
\end{small}\\
\begin{small}
$^4$Dipartimento di Fisica, Universit\'a di Pisa, Largo B. Pontecorvo 3, 56127 Pisa, Italy.
\end{small}
}

\date{}

\maketitle

\abstract{Local solutions are a vivid and long-living topic in gravitational physics. We consider exact static pseudo-spherically symmetric solutions of
semi-classical Einstein's equations in the presence of the trace anomaly contribution. We investigate black hole solutions and propose new metrics describing traversable wormholes. Thanks to the trace anomaly, wormholes are realized in vacuum and, nevertheless, violate the null energy condition.}
\section{Introduction}

The local solutions represent a test bench for any gravity theory and 
have drawn the attention of physicists for over a hundred years.
Exact vacuum solutions of General Relativity (GR) are well known, while in the non-vacuum case, we can reproduce a wide class of different metrics with
interesting features. See for instance the case of regular black holes \cite{Hayward} and the case of wormholes \cite{Morris:1988cz, Damour}.
Other results can be obtained in alternative theories of GR: for example, one may consider the presence of new fields in the underlying theory \cite{Motohashi:2018wdq, Filippini:2017kov} or
modify the gravitational Lagrangian as in $F(R)$-gravity or in other modified theories of gravity \cite{Calza:2018ohl,Kalita:2021hyt,Lobo:2011fr,Ghosh:2021msy,Oliveira:2011vu,Bahamonde:2021srr,Kehagias:2014sda}. In those cases, finding analytical exact solutions appears as a formidable task since the equations of motion are more complex with respect to GR ones. For this reason, the numerical approach has also been pursued \cite{Sullivan:2019vyi}.

The conformal anomaly \cite{Duff}, or trace anomaly (TA), is a quantum-level anomaly breaking a conformal symmetry of the classical theory. It plays an important role in quantum field theories in curved space-time, and it is object of various studies since many decades \cite{Capper:1974ic, Duff:1977ay, BrownLowell, BrownLowell2, Fradkin:1983tg, OdiSha, Antoniadis:1992xu, Antoniadis:1992ep, Bastianelli:1992ct} (see also the recent works \cite{Ferrara:2020zef, Duff:2020dqb}).
For example, the accelerated expansion in the early universe, i.e., inflation \cite{Guth, Sato}, can be realized by quantum effects of conformally invariant fields \cite{in1, in2, in3}, and various astrophysical and cosmological applications exist related to the quasi thermal emission of black-holes \cite{Christensen:1977jc} and the cosmological constant problem \cite{Robinson:2005pd,Iso:2006wa}. The trace anomaly also plays a role in string theory and statistical mechanics.

In a homogeneous and isotropic universe, the trace anomaly totally determines its energy-momentum tensor. For example, some relevant results have been found by implementing conformal anomaly in different inflationary models \cite{FRW1, FRW2,Nojiri:2017ncd, FRW3, FRW4, FRW5, FRW6, FRW7}. However, in four-dimensional spherically symmetric static (SSS) space-time, the trace anomaly energy-momentum tensor is known up to an arbitrary function of position \cite{Cai1} and usually we can consider only perturbative solutions around the background Einstein's equations. 

An analytical SSS black hole (BH) solution has been found in Refs. \cite{Cai1}, where the so-called ``Type A'' anomaly has been studied up to one loop quantum corrections, and the logarithmic correction to the black-holes entropy has been computed. Moreover, in Ref. \cite{Cai2}, a second BH solution has been derived in Anti-De Sitter (AdS) space-time by considering a negative cosmological constant contribution. 

In this paper, similarly to the seminal works of Cai {\textit{et al.}} in \cite{Cai1, Cai2}, we analyze pseudo-spherically symmetric static (SSS) solutions of GR in the presence of trace anomaly. We will maintain the focus on SSS space-time in the presence of one loop trace anomaly and we enlarge the discussion for generic topological cases. We reconsider BH solutions in vacuum and non-vacuum cases and we derive for wormhole (WH) solutions, both of Type-A case and not. We propose two new solutions, showing that traversable wormholes can be reproduced by considering quantum effects without the presence of exotic matter. 

The paper is organized as follows. In Sec. {\bf 2} we introduce the equations of motion of GR with trace anomaly on topological SSS space-time. In Sec. {\bf 3} we reconsider BH solutions in vacuum and non-vacuum cases. Sec. {\bf 4} is devoted to WH metrics, here we derive and discuss two new solutions. Conclusions and final remarks are given in Sec. {\bf 5}.

In our convention, the speed of light $c = 1$ and the Planck mass $1/8\pi G_N = 1$. We also adopt the "mostly plus" metric convention.

\section{Trace anomaly}

The trace anomaly is a quantum effect present when we consider the contribution of (mass-less) matter fields. In the Standard Model of particle physics, almost a hundred among mass-less scalar fields, Dirac spinors, and vectors are considered, and this number may be doubled if the Standard Model is embedded in a supersymmetric theory. Moreover, the number of scalar fields may skyrocket up to $10^5$ if one considers axion-like particles coming from string compactification \cite{Arvanitaki:2009fg, March-Russell:2021zfq, Calza:2021czr}. Fields action in curved space-time is conformal invariant, but a renormalization must be introduced to cancel the divergences from the one-loop vacuum contributions. As a consequence, the counter-terms necessary to avoid the poles of the divergences break the conformal invariance of the matter action itself. From a classical point of view, the trace of the energy-momentum tensor in a conformally invariant theory vanishes, but after the renormalization procedure we get a non-zero trace of an anomalous energy-momentum tensor, what is called quantum anomaly or trace anomaly \cite{an1, an2, an3, an4}. 

If we include the backreaction of the quantum fields to a curved space-time geometry in the field equations of GR we obtain 
\begin{equation}
R_{\mu\nu}  -\frac{1}{2}R g_{\mu\nu}=\left<T_{\mu\nu}\right>\,,
\label{feq}
\end{equation}
where: $R_{\mu\nu}$ and $R$ are the Ricci tensor and scalar, respectively; $g_{\mu\nu}$ is the metric and $\left<T_{\mu\nu}\right>$ is an effective energy-momentum tensor by quantum loops
and must be covariantly conserved, namely it obeys the continuity equation $\nabla^\mu T_{\mu\nu}=0$.
In general, it is almost impossible to derive the correct expression of such a tensor for a generic space-time, and some assumptions on the background must be considered. On the other hand, in four dimensions, the trace anomaly has the following form, 
\begin{equation}
<T>=\lambda C^2-\alpha G\,,\label{traceanomaly}
\end{equation}
where: 
$T=g^{\mu\nu}T_{\mu\nu}$ is the trace of the energy-momentum tensor; $\lambda$ and $\alpha$ are two positive constants depending on the degrees of freedom of quantum fields. The scalars $G$ and $C^2$ are the Gauss-Bonnet four-dimensional topological invariant and
the square of the Weyl tensor, namely 
\begin{equation}
G=R^2-4 R_{\mu\nu}R^{\mu\nu} +  R_{\mu\nu\sigma\xi}R^{\mu\nu\sigma\xi}\,,
\quad
C^2\equiv C_{\mu\nu\sigma\xi}C^{\mu\nu\sigma\xi}=\frac{R^2}{3}-2 R_{\mu\nu}R^{\mu\nu} +  R_{\mu\nu\sigma\xi}R^{\mu\nu\sigma\xi}\,,
\end{equation}
where $R_{\mu\nu\sigma\xi}$ is the Riemann tensor, and correspond to the so-called ``Type A'' and ``Type B'' anomaly, respectively \cite{type}.

We are interested in the (pseudo-)spherically symmetric static metrics 
\begin{equation}
ds^2=-f(r)dt^2+\frac{1}{g(r)} dr^2+r^2\,\left(\frac{d\rho^2}{1-k\rho^2}+\rho^2 d\phi^2\right) \,,\label{metric}
\end{equation}
where $f\equiv f(r)$ and $g\equiv g(r)$ are functions of the radial coordinate $r$ only, and the manifold topology will be either a sphere, a torus, or a compact hyperbolic manifold, according to whether $k=1,0,-1$, respectively. Thus, the Gauss-Bonnet reads,
\begin{equation}
G= 
\frac{-2 k f f' g'+2g\left(-2kf f'' +3f f' g' +kf'^2\right)-2g^2\left(f'^2-2f f''\right)}{r^2f^2}\,,
\label{G}
\end{equation}
while the squared Weyl tensor assumes the form,
\begin{equation}
    C^2=\frac{1}{12}\left\lbrace\frac{r^2 g f'^2 + 2 f^2(2k-2g+rg')-rf[rf'g'-2g(f'-rf'')]}{f^2 r^2}\right\rbrace^2\,.
\end{equation}
In the equations above the prime index denotes the derivative with respect to the radial coordinate $r$.
The symmetry of the metric brings to the identification
\begin{equation}
T_0^0=-\rho\,,  \quad
T_1^1=p\,,\quad T_2^2=T_3^3=p_\perp\,,
\end{equation}
where $\rho$ and $p$ are the effective energy density and radial pressure of trace anomaly, while $p_\perp$ is its effective transversal pressure.

The first two field equations in \eqref{feq}, with the metric \eqref{metric}, read \cite{Chinaglia}
\begin{align}
\frac{d}{dr}\left[r(k-g)\right]&=8\pi r^2\rho\,, \label{00}\\
\frac{g}{f r}f'+\frac{(g-k)}{r^2}&=8\pi p\,. \label{11}   
\end{align}
The above equations with the conservation law read
\begin{equation}
4f(p-p_\perp)+(\rho+p)r f'+2r p' f=0\,, 
\label{conslaw}
\end{equation}
and considering the constraint in (\ref{traceanomaly}) we obtain
\begin{equation}
-\rho+p+2p_\perp=\lambda C^2 -\alpha G\,,
\label{trace}
\end{equation}
which can be rewritten as
\begin{equation}
2r f p'+r f'(\rho+p) +2f(3p-\rho)= 2(\lambda C^2-\alpha G)f\,. \label{last}
\end{equation}
In principle, given an Equation of State (EoS) $p=p(\rho)$, we get a system of differential equations (\ref{00}), (\ref{11}) and (\ref{last}) with respect to the quantities $p\,,f\,,g$, while $p_\perp$ follows from (\ref{trace}). In the next sections, we solve this differential equations system with some metric Ansatz.
Firstly, we revisit the black hole case introduced in \cite{Cai1, Cai2}, and then we analyze wormhole solutions.

\section{Black hole solutions}

Following \cite{Cai1}, as a test of the procedure, we firstly consider black hole solutions in the presence of Type A anomaly only, when $\lambda=0$ and the contribution of the squared Weyl tensor disappears. The metric (\ref{metric}) may describe a black hole whose horizon is located at the zeros of the metric functions $g(r)$ and $f(r)$, simultaneously. Since the conditions $g(r)=0$ and $f(r)=0$ must be satisfied for the same value of $r=r_H$, a direct consequence of the field equations (\ref{00})--(\ref{11}) is that $p=-\rho$ on the BH horizon $r_H$. By generalizing this condition to the whole space-time, we can rewrite (\ref{last}) as
\begin{equation}
4p+r p'=-\alpha G\,.   
\end{equation}
Moreover, with the EoS $p=-\rho$ we are forced to put $f(r)=g(r)$ in order to solve (\ref{00}) and (\ref{11}) together. We obtain
\begin{equation}
p=\frac{2\alpha g(r)^2-4k\alpha g(r)+4r\alpha(k-g)g'(r)}{r^4}-\frac{c_1}{r^4}\,, 
\label{rad0}
\end{equation}
where $c_1$ is an integration constant and we used (\ref{G}). The field equations (\ref{00})--(\ref{11}) finally lead to
\begin{equation}
g(r)=k-\frac{r^2}{32\pi\alpha}
\pm\frac{r^2}{32\pi\alpha}
\sqrt{1+\frac{512\pi^2\alpha}{r^4}( 2 k^2 \alpha+c_1)-\frac{c_0}{r^3}}\,,
\label{Cai0}
\end{equation}
where $c_0$ is a constant.
The form of the effective transversal pressure $p_\perp$ can be easily computed by (\ref{trace})
\begin{equation}
    p_\perp=\frac{c_1 - \left( 2 \alpha g(r) (-2 k +g(r)-2 r g'(r) + r^2 g''(r)) + g'(r)(4 k r \alpha +2 \alpha r^2 g'(r)) -2 k r^2 \alpha g''(r) \right)}{r^4} 
    \,.
\end{equation}
The solution (\ref{Cai0}) is the topological generalization of the one found in \cite{Cai1} for $k=1$. We have two branches, corresponding to the signs ``$\pm$'' in (\ref{Cai0}). For the branch with plus sign in the spherical case $k=1$, we obtain the Reissner-Nordstr{\"o}m solution for large values of $r$, namely
\begin{equation}
g(r)\simeq 1-\frac{c_0}{64\pi \alpha r}
+\frac{8\pi}{r^2}(2 \alpha+c_1)\,.
\end{equation}
If $c_0>0$ the solution describes a static black hole. Thus, $c_0$ corresponds to the mass of the BH yielding logarithmic corrections to the standard area law of the entropy \cite{Cai1}. On the other hand, by choosing the minus sign in (\ref{Cai0}), for large values of $r$ we have
\begin{equation}
g(r)\simeq k-\frac{c_0}{64\pi \alpha r}
+\frac{8\pi}{r^2}(2k^2\alpha+c_1)-\frac{r^2}{16\pi\alpha}\,,
\end{equation}
i.e., an asymptotically De Sitter space-time. The metric describes a BH with various topologies and $c_0>0$. In particular, we observe that for the flat and hyperbolic cases with $k=0, -1$, an event horizon appears as soon as $(2k^2\alpha+c_1)>0$ (for quasi topological Reissner Nordstr{\"o}m solution see also \cite{Mann}).  
\\
\\
In principle, some other analytical BH solutions with $p=-\rho$ (and therefore $f(r)=g(r)$)
can also be found by considering the full contribution from the trace anomaly, but in general, we have to introduce a constraint on the parameters $\alpha$ and $\lambda$. This procedure seems unnatural, since $\alpha$ and $\lambda$ depend on the degrees of freedom of quantum fields.
 
As an example, we consider the topological Schwarzschild dS/AdS solution
\begin{equation}
f(r)=g(r)=k-\frac{c_0}{r}+{c_1}{r^2}\,,  
\end{equation}
where $c_{0,1}$ are constants. 
In order to obtain $g_{00}>0$ when $r\rightarrow 0^+$, we assume $c_0>0$. In the spherical topological case, both a cosmological horizon and an event horizon appear. On the other hand, if $k=0, -1$, we need an asymptotically AdS space-time with $c_1>0$ and in this case, an event horizon should be present.
Since
\begin{equation}
p=-\rho=\frac{c_1}{8\pi }\,,\quad p_\perp=0\,,   
\end{equation}
it turns out that (\ref{last}) is satisfied by making the choice,
\begin{equation}
\lambda = \alpha\,, \quad c_1=-\frac{1}{16 \pi \alpha}\,,
\end{equation}
or
\begin{equation}
\lambda = \alpha\,, \quad c_1= 0\,.
\end{equation}
Therefore the spherical case $k=1$ is the only relevant one. We note that $c_1$ is fixed by the model, such that $c_0$ can be correctly identified with the BH mass.
\\ \\
To obtain BH solutions without constraints on $\alpha$ and $\lambda$, we may introduce additional fluid contributions to the field equations \cite{Saha}. In what follows, we will denote with the subscript ``$F$'' the contributions of the fluid, and with the subscript 
``$TA$'' is the contribution of the trace anomaly.
The effective energy density and pressure in (\ref{00})--(\ref{11}) and (\ref{conslaw}) are
\begin{equation}
\rho=\rho_{TA}+\rho_F\,,\quad p=p_{TA}+p_{F}\,,    \quad p_\perp=p_{\perp TA}+p_{\perp F}\,,
\end{equation}
and (\ref{last}) is
\begin{equation}
2r f p'+r f'(\rho+p)
+2f(3p-\rho)= 2f(\lambda C^2-\alpha G)+2f(-\rho_F+p_F+2p_{\perp F})\,.
\label{last2}
\end{equation}
Firstly, we recover the result of \cite{Cai2} by taking into account the contribution of a cosmological constant, namely
\begin{equation}
\rho_F=\Lambda\,,\quad  p_F=p_{\perp F} =-\Lambda\,,
\end{equation}
where $\Lambda$ is a constant. In the presence of Type A anomaly and by assuming $p=-\rho$ we find
\begin{equation}
 f(r)=g(r)= 
 k-\frac{r^2}{32\pi\alpha}
 \pm\frac{r^2}{32\pi\alpha}\sqrt{1+\frac{64 \pi  \alpha  \left(c_1+16 \pi  \alpha  k^2\right)}{r^4}+\frac{1024 \pi ^2
   \alpha ^2 c_0}{r^3}-\frac{512}{3} \pi ^2 \alpha  \Lambda}\,,
\end{equation}
where $c_{0,1}$ are constants. This is the topological generalization of the BH solution found in \cite{Cai2}, where various aspects of the metric are discussed. 

Moreover, many BH solutions can be reconstructed by assuming different fluid contributions. We consider the case of a perfect fluid in the presence of the full contribution of the trace anomaly ($\alpha, \lambda\neq 0$). The stress-energy tensor of a perfect fluid is given by
\begin{equation}
T^\mu_{\nu F} =diag (-\rho_F, p_F, p_F, p_F)\,, 
\end{equation}
namely $p_{\perp F}=p_F$.
The barotropic Equation of State reads
\begin{equation}
p_F=\omega \rho_F\,,    
\end{equation}
with $\omega$ constant. If one looks for the topological Reissner-N{\"o}rdstrom solution
\begin{equation}
f(r)=g(r)=k-\frac{c_0}{r}+\frac{c_1}{r^2}\,,    
\end{equation}
with $c_0\,,c_1$ constants, we find the effective energy density and pressure
\begin{equation}
\rho= \frac{c_1}{8\pi r^4}\,,\quad 
p=- \frac{c_1}{8\pi r^4}\,,
\end{equation}
and (\ref{last2}) states that we need a perfect fluid with $\omega\neq 1/3$ (non trace-less fluid). On shell we get
\begin{equation}
\rho_F=\frac{8 c_1^2 (5\alpha-6\lambda)-48 c_0 c_1 r(\alpha-\lambda)+12c_0^2 r^2(\alpha-\lambda)}{r^8(3\omega-1)}\,.    \end{equation}
In the next section, we investigate wormhole vacuum solutions in the presence of the quantum anomaly.

\section{Wormhole solutions}
%%%%%%%%%%%%%%%%%%%%%%%%%%%%%%%%%%%%%%%%%%%%%
In this section, we study some wormhole solutions. The traversable wormholes are realized in the framework of GR only in the presence of matter sources violating the null energy condition \cite{WH,WH2,WH3,WH4,WH5,WH6,WH7,WH8,WH9,WH10,WH11}. In modified gravity, such a kind of solutions can be obtained also in vacuum since the role of the anti-gravitational matter can be played by the modification of gravity itself \cite{Calza:2018ohl,WH8,WH9,WHMG,WHMG4}. In this paper, we work in the GR framework and we aim to identify the trace anomaly as the source of the negative pressure fluid violating the null energy condition. In such a way, the trace anomaly prevents the wormhole throat from pinching off.
For our purpose we replace the metric function $f(r)$ in (\ref{metric}) with 
\begin{equation}
f(r)=\text{e}^{2\Phi(r)}\,,
\end{equation}
where $\Phi\equiv\Phi(r)$ is a new redshift metric function, such that the resulting line element reads
\begin{equation}
ds^2=-\text{e}^{2\Phi(r)}dt^2+\frac{1}{g(r)} dr^2+r^2\,\left(\frac{d\rho^2}{1-k\rho^2}+\rho^2 d\phi^2\right) \,.\label{metric2}
\end{equation}
A traversable wormhole occurs if the radial coordinate is embedded by a minimal radius at $r=r_0$, the so-called ``throat'', and no horizons appear. In particular, the function $\Phi(r)$ must be finite and regular everywhere along the throat and the following conditions must be satisfied \cite{WHcond,WHcond2,WHcond3}:
\begin{itemize}
\item $\Phi_\pm(r)$ and $g_\pm(r)$ are well defined smooth functions for all $r \geq r_0$;
\item  $\Phi'_+(r_0)=\Phi'_-(r_0)$; 
\item $g_\pm(r_0)=0$ and  $g_\pm(r)>0$ for all $ r \geq r_0$;
\item $g'_+(r_0)=g'_-(r_0)>0$.
\end{itemize}

Since the proper distance is defined as
\begin{equation}
l(r)=\pm\int ^r_{r_0}\frac{d\tilde r}{g(\tilde r)} \,,
\end{equation}
its minimal value is reached for $r=r_0$, while the positive and negative values of $l\equiv l(r)$ correspond to the lower and upper universes connected through the throat of the wormhole. The travelling time necessary to cross the wormhole between $l(r_1)=-l_1<0$ and $l(r_2)=+l_2>0$
is given by,
\begin{equation}
\Delta t=\int_{l_1}^{l_2}   \frac{d l}{v \text{e}^{\Phi(l)}} \,,
\end{equation}
where $v=d l/[\text{e}^{\Phi(l)} dt]$ is the radial velocity of the traveler as he/she passes radius $r$ \cite{Morris:1988cz}. Thus, if $\Phi'(r)<0$ the repulsive tidal force makes it impossible to travel across the throat.

Firstly, we look for wormhole solutions with
\begin{equation}
\Phi(r)=0\,,    \label{ex0}
\end{equation}
namely, we consider a vanishing tidal force where the proper time measured by a static observer corresponds to the coordinate time $t$ (the result can be generalized to the case $\Phi(r)=\text{const}$).
This choice greatly simplifies equation (\ref{last}) if we take into account the Type A anomaly ($\lambda=0$), since the Gauss-Bonnet term vanishes. By using (\ref{00}) and (\ref{11}) we obtain
\begin{equation}
g+r g' = k\,,
\end{equation}
whose solution is given by
\begin{equation}\label{sol1}
g(r)=k-\frac{c_0}{r}\,,    
\end{equation}
where $c_0$ is an integration constant. 
The field equations (\ref{00})--(\ref{11}) lead to
\begin{equation}
\rho=0\,, \quad p=-\frac{c_0}{8\pi r^{3}}\,,\label{pWH0}
\end{equation}
and the metric finally reads
 \begin{equation}
ds^2=-dt^2+\frac{1}{\left(k-\frac{c_0}{r}\right)} dr^2+r^2\,\left(\frac{d\rho^2}{1-k\rho^2}+\rho^2 d\phi^2\right) \,.\label{metric3}
\end{equation}
Note that the metric does not depend on the parameter $\alpha$. The trace anomaly supports the solution through an effective pressure, but on shell, its trace contribution disappears.
Wormhole solutions can be obtained for the spherical case with $k=1$ and $c_0>0$. In this case, the radius of the throat corresponds to $r=c_0$.
The Gauss-Bonnet (and therefore the Ricci scalar) vanishes, and the trace anomaly corresponds to a (trace-less) effective fluid with zero energy density and radial (negative)  pressure (\ref{pWH0}).
The transversal pressure can be derived from (\ref{trace}) and simply reads,
\begin{equation}
p_T=-\frac{p}{2}=   \frac{c_0}{16\pi r^3}\,. 
\end{equation}
Wormhole solutions require the violation of the Null Energy Condition (NEC) near the throat \cite{Visser:2003yf,Visser,Kanti:2011jz, Barcelo:2000zf}, which is usually written in terms of the stress-energy tensor as 
\begin{equation}
T_{\mu\nu}n^\mu n^\nu \geq 0\,, \quad n_\mu n^\mu =-1\,,    
\end{equation}
where $n^\mu$ is a null-vector and $T_{\mu \nu}= diag(\rho, p,p,p)$.
Among the different energy conditions, the NEC is particularly interesting since it is the weakest: if violated, all the other energy conditions are consequently violated\footnote{ 
The energy conditions read:
\begin{itemize}
    \item Null Energy Condition (NEC) $\rho + p > 0$
    ;\item Weak Energy Condition (WEC) $\rho + p > 0 $ , $\rho>0$;
    \item Dominant Energy Condition (DEC) $\rho> |p|$
    ;\item Strong Energy Condition (SEC) $\rho + p>0$ , $\rho+ 3p >0$.
\end{itemize}}.
In our case, the NEC is clearly violated as a consequence of (\ref{pWH0}). Thus, the conformal anomaly plays the role of an exotic fluid supporting the WH solution.

Consider then the class of traversable WH models with
\begin{equation}
\Phi(r)=\frac{1}{2}\log\left[\left(\frac{r}{r_0}\right)^z\right]\,,\quad z>0\,,  
\end{equation}
where $z$ is a constant parameter.
A choice that greatly simplifies the form of the Gauss-Bonnet and of the square of the Weyl tensor is given by $z=2$. In this case, the metric has the form
\begin{equation}
ds^2=- r^2 dt^2+\frac{1}{g(r)} dr^2+r^2\,\left(\frac{d\rho^2}{1-k\rho^2}+\rho^2 d\phi^2\right) \,.\label{metric4}
\end{equation} 
Equation (\ref{last}) leads to the solution
\begin{equation}
    g(r)=
    \frac{k}{3}-\frac{r^2}{32\pi\alpha}\pm
    \frac{\sqrt{\xi(r)}}{96 \pi  \alpha }\,,
    \label{gWH2}
\end{equation}
where
\begin{equation}
\xi(r)=1024 \pi ^2 \alpha ^2 \left(9 c_0+k^2\right)+2048 \pi ^2 \alpha  k^2  \lambda  \log (r)+9 r^4 \,,\label{gWH2bis}
\end{equation}
with $c_0$ constant. The solution has two branches, corresponding to the $\pm$ signs.
The relevant case that reproduces a traversable wormhole 
corresponds to the plus sign. Thus, 
for large values of $r$ we get,
\begin{equation}
g(r)\simeq \frac{k}{3}\,,  
\end{equation}
and the spherical case with $k=1$ preserves the metric signature. The throat is located at $r=r_0$, such that $g(r_0)=0$. In order to have $g'(r_0)>0$ we must require
\begin{equation}
\frac{2(3r_0^2+16\pi\lambda)}{r_0(9r_0^2-96\pi \alpha)}>0\,.\label{gprime}  
\end{equation}
If we restrict the analysis to the Type A anomaly, we can find an analytical value for the radius of the throat. Since $g(r_0) = 0$ and therefore $\xi(r_0)= \left(3 r_0^2 -32 \pi  \alpha \right)^2$ we get
\begin{equation}
    c_0+\frac{r_0^2}{48 \pi  \alpha }=0\,,
\end{equation}
whose solution is
\begin{equation}
r_0=4\sqrt{-3\pi\alpha c_0}\,.
\label{lambda0}
\end{equation}
The condition $g'(r_0)>0$ simply reads $r_0>\sqrt{\frac{32}{3} \pi \alpha}$ implying
\begin{equation}
 \quad c_0<-\frac{2}{9}\,.
\label{Condc0}
\end{equation}
An implicit solution can be found also for the general case with $\lambda \neq 0$. Again, 
$\xi(r_0)= \left(3 r_0^2 -32 \pi  \alpha \right)^2$ implying 
\begin{equation}
    c_0+\frac{r_0^2}{48 \pi  \alpha }+\frac{2 \lambda  \log (r_0)}{9 \alpha }=0\,.
\end{equation}
As a consequence, the radius of the throat is
\begin{equation}
    r_0 = 4  \sqrt{\frac{\pi \lambda}{3} W\left[ \frac{3 }{16 \pi  \lambda }e^{-\frac{9 \alpha  c_0}{\lambda }}\right]}
\,,\label{r0imp}
\end{equation}
where $W[x]$ is the principal solution of the (multi-valued special complex) Lambert function. If we restrict to real values the Lambert function is the inverse function of $x=W e^W $. 
%the same way the square root or the logarithm allow to invert the of elevation to power or exponential, namely:
%\begin{itemize}
   % \item $y^2=x \rightarrow y= \pm \sqrt{x}$, with $x \in [0,\infty)$
%    \item $e^y=x \rightarrow y= \log(x)$ with $x \in (0,\infty)$
 %   \item $y e^y=x \rightarrow y= W(x)$ with $x \in (-\frac{1}{e},\infty)$
%\end{itemize}

In the limit $\lambda\rightarrow 0^+$ we obtain $ \sqrt{ \lambda W\left[ \frac{3 }{16 \pi  \lambda }e^{-\frac{9 \alpha  c_0}{\lambda }} \right]} = 3 \sqrt{-\alpha c_0}$ and the result in (\ref{lambda0}) is restored.

The condition $g'(r_0)>0$ implies
\begin{equation}
    \frac{\sqrt{\lambda } \left(W\left[\frac{3 e^{-\frac{9 \alpha  c(0)}{\lambda }}}{16 \pi  \lambda }\right]+1\right)}{2 \sqrt{3 \pi } \sqrt{W\left[\frac{3 e^{-\frac{9
   \alpha  c(0)}{\lambda }}}{16 \pi  \lambda }\right]} \left(\lambda  W\left[\frac{3 e^{-\frac{9 \alpha  c(0)}{\lambda }}}{16 \pi  \lambda }\right]-2 \alpha \right)}>0\,,
\end{equation}
which becomes (\ref{gprime}) by using (\ref{r0imp}).
Taking into account that $\lambda>0$, both the numerator and $2 \sqrt{3 \pi } \sqrt{W\left[\frac{3 e^{-\frac{9
   \alpha  c(0)}{\lambda }}}{16 \pi  \lambda }\right]}$ are positive quantities, such that the former inequality reduces to
   \begin{equation}
    \lambda  W\left[\frac{3 e^{-\frac{9 \alpha  c_0}{\lambda }}}{16 \pi  \lambda }\right]>2 \alpha   \,.
   \end{equation}
Since $\alpha>0$ we obtain
   \begin{equation}
       c_0<-\frac{\lambda  \log \left(\frac{32}{3} \pi  \alpha  e^{\frac{2 \alpha }{\lambda }}\right)}{9 \alpha }\,.
   \end{equation}
Note that
 \begin{equation}
     \lim_{\lambda \rightarrow 0^+}{\left( -\frac{\lambda  \log \left(\frac{32}{3} \pi  \alpha  e^{\frac{2 \alpha }{\lambda }}\right)}{9 \alpha } \right) }= -\frac{2}{9}\,,
 \end{equation}
and we obtain the condition in (\ref{Condc0}).
 
The explicit forms of $\rho\,,p$ and $p_\perp$ with $\alpha\,,\lambda\neq 0$ are derived as
\begin{align}
    \rho &=\frac{1}{12 \pi  r^2} +\frac{3}{256 \pi ^2 \alpha }-\frac{3 r^2}{128 \pi ^2 \alpha  \sqrt{\xi(r)}}-\frac{\sqrt{\xi(r)}}{768 \pi ^2 \alpha  r^2}-\frac{4 \lambda }{3 \sqrt{\xi(r)} r^2}\,,\nonumber\\
     p&=\frac {1} {256 \pi^2 \alpha}\left(\frac { 
   \sqrt{\xi(r)}} { r^2} -3\right)\,,\nonumber\\
   p_\perp&=\frac{\alpha  \left(12 c_0+\frac{4}{3}\right)}{\sqrt{\xi } r^2}+\frac{3 \left(\frac{3 r^2}{\sqrt{\xi }}-1\right)}{256 \pi ^2 \alpha }+\frac{4 \lambda  (2 \log
   (r)+1)}{3 \sqrt{\xi } r^2}+\frac{1}{24 \pi  r^2}\,.
\end{align}
The NEC on the throat reads 
\begin{equation}
(\rho+p)\lvert_{r=r_0}  =
-\frac{16 \pi\lambda+3r_0^2}{4 \pi  r_0^2 \left(9 r_0^2-96 \pi ^2 \alpha \right)}<0\,,
\end{equation}
which is clearly violated as a consequence of (\ref{gprime}).

As a final remark about the choice of the minus sign in (\ref{gWH2})--(\ref{gWH2bis}), for large values of $r$ we obtain
\begin{equation}
g(r)\simeq \frac{k}{3}-\frac{2r^2}{32\pi\alpha}\,,   
\end{equation}
with the appearance of a cosmological horizon when $k=1$.

\section{Conclusions}
In this paper, we studied exact analytic static pseudo-spherically symmetric solutions in semi-classical Einstein's equations in the presence of trace anomaly. These solutions are characterized by a fluid-like component that originates from the trace anomaly. The solutions we analyzed are suitable to describe compact objects such as black holes and wormholes.
The space of solutions is very large and provides a rich phenomenology. Nevertheless, finding exact solutions is not a trivial task since it is not possible to reconstruct the exact form of effective stress-energy tensor associated with the trace anomaly and some assumptions are necessary.
This difficulty led previous studies to tackle this problem from different perspectives \cite{Ho:2018fwq,Kawai:2014afa,Kawai:2021qdk} and may lead future studies to consider numerical approaches to enlarge the space of solutions.

Regarding black hole solutions, as previous studies have already shown \cite{Cai1, Cai2}, the metric is in general different from Schwarzschild.
This calls for a comparison with the Solar System tests, applying the Post Newtonian Parameters formalism in order to constrain the parameter space. Gravitational waves from black hole mergers may provide additional information and help to further constrain the space of parameters. In the case of Type A anomaly (namely $\lambda=0$) we generalized the solution in \cite{Cai1} to hyperbolic and flat topological cases. This solution does not pose any constraint on $\alpha$ which can be determined by the underlying theory of particle physics, and in the limit of large $r$ in the case $k = 1$ the solution reduces to the Reissner-Nordstr\"{o}m solution. In the case of a full contribution of the trace anomaly, we recover notable solutions like the topological Schwarzschild dS/AdS ones only if we introduce specific constraints on the parameters $\alpha$ and $\lambda$, otherwise, it is necessary to consider the non-vacuum case. 
The introduction of a cosmological constant allows us to recover and generalize the result in \cite{Cai2}, while in the presence of perfect fluids, it is possible to find the Reissner-Nordstr\"{o}m solution.

In the case of wormholes, we found two new traversable solutions. These solutions show a striking conceptual advantage with respect to GR: they do not require the presence of exotic matter sources violating GR energy conditions. This unpleasant role is played by the trace anomaly contribution itself.  The first solution is obtained for a Type A anomaly and represents one of the simplest wormhole solutions. The latter solution is more general and obtained considering the full trace-anomaly contribution. In this case, the time-time component of the metric is Lifschitz-like inspired and the solution features a more complex description.

\section*{Acknowledgements}

M.C. is supported by the FCT doctoral grant SFRH/BD/146700/2019.

\end{document}